\documentclass[doublecol]{epl2} 
\usepackage[T1]{fontenc}
\usepackage[latin9]{inputenc}
\usepackage{color}
\usepackage{amsmath}
\usepackage{amssymb}
\usepackage{graphicx}
\usepackage{esint}

\title{Thermomagnonic spin transfer and Peltier effects in insulating magnets}

\author{Alexey A. Kovalev \inst{1} \and Yaroslav Tserkovnyak \inst{2}}

\institute{                    
  \inst{1}Department of Physics and Astronomy, University of California, Riverside, California 92521, USA

  \inst{2} Department of Physics and Astronomy, University of California, Los Angeles, California 90095, USA
}

\abstract{
We study the coupled magnon energy transport and collective magnetization dynamics in ferromagnets
with magnetic textures. By constructing a phenomenological theory
based on irreversible thermodynamics, we describe motion of domain
walls by thermal gradients and generation of heat flows by magnetization
dynamics. From microscopic description based on magnon kinetics, we estimate the transport coefficients
and analyze the feasibility of energy-related applications in insulating ferromagnets, such as
yttrium iron garnet and europium oxide.}

\begin{document}

\maketitle

\section{Introduction} 
 Most electronic devices rely on charge current
flows controlled by applied voltages. It has been realized that spin-polarized charge
flows can also be induced by fictitious electromagnetic fields due
to magnetic texture dynamics \cite{Volovik:mar1987, Tatara:nov2008}. Recently, the
possibility of employing spin in electronic logic devices has also been suggested \cite{Dery:may2007}. However, since it may be energy-costly to create
and maintain spin imbalances electrically, the fictitious electromagnetic
fields induced by magnetic textures may become useful for such applications.
Reciprocal motion of magnetic textures due to weak charge-current-induced spin transfer
is governed by dissipative torques (so-called
``$\beta$ terms") \cite{Zhang:sep2004,Thiaville:feb2005,Tserkovnyak:Oct2006,Kohno:nov2006,Duine:jun2007}
associated with microscopic spin misalignments, which will also play an important role in our study.

The spin flows are known to coexist with heat flows
in transition metals, where both fluxes can be effectively driven by
the moving magnetic texture \cite{Kovalev:Sep2009,Kovalev:nov2010}. Similar effects
have also been studied in magnetic semiconductors \cite{Hals:2010}.
The reciprocal action of the spin-transfer torque (STT) on the magnetization
\cite{Slonczewski:1996} is extremely important for applications,
e.g., spin-transfer torque memory and nonvolatile logic. Spin torques due to pure
spin currents have been observed in experiments on spin pumping
by magnetic precession \cite{Heinrich:May2003} or spin Hall effect along magnetic insulator surface \cite{Kajiwara:mar2010}. Thermal spin torques
have been studied in magnetic nanopillars \cite{Hatami:aug2007}
with recent developments suggesting that they can be more efficient compared
to electrically generated spin torques \cite{Slonczewski:Aug2010}.

Spin and heat currents can be also induced in insulating systems,
such as yttrium iron garnet (YIG), by the external microwave
magnetic field \cite{Kajiwara:mar2010}. Alternatively, the temperature
gradient can lead to a spin imbalance by the spin Seebeck effect \cite{Uchida:nov2010}.
The heat currents accompanying the spin-wave (magnon) flows in an
insulating ferromagnet $\mbox{Lu}_{2}\mbox{V}_{2}\mbox{O}_{7}$ have
been used to detect the magnon Hall effect \cite{Onose:2010}. Here, we are interested in
the possibility to use fictitious electromagnetic fields induced by magnetic textures
to control magnon spin flows in a manner similar to spin-polarized charge currents.

In this Letter, we study interplay between magnon-carried heat and spin currents
and magnetic texture dynamics in ferromagnets by formulating a hydrodynamic
phenomenological description of magnonic and thermal currents coupled to geometric gauge fields (the charge currents are ignored in our description but can
be readily reintroduced in case of metals \cite{Kovalev:Sep2009}). By considering
dissipative corrections (``$\beta$ terms"), we show that they play an important role as they do in electronic systems,
enabling domain-wall (DW) motion below the Walker breakdown and heat
pumping \cite{Kovalev:Sep2009}. To justify our phenomenology, we
also formulate a ground-up kinetic description of thermal magnons, which
allows us to identify the phenomenological coefficients entering
the hydrodynamic theory. The DW motion and heat pumping
are studied in YIG and EuO and the feasibility of energy-related applications
is discussed.

\begin{figure}
\onefigure[width=0.9\linewidth]{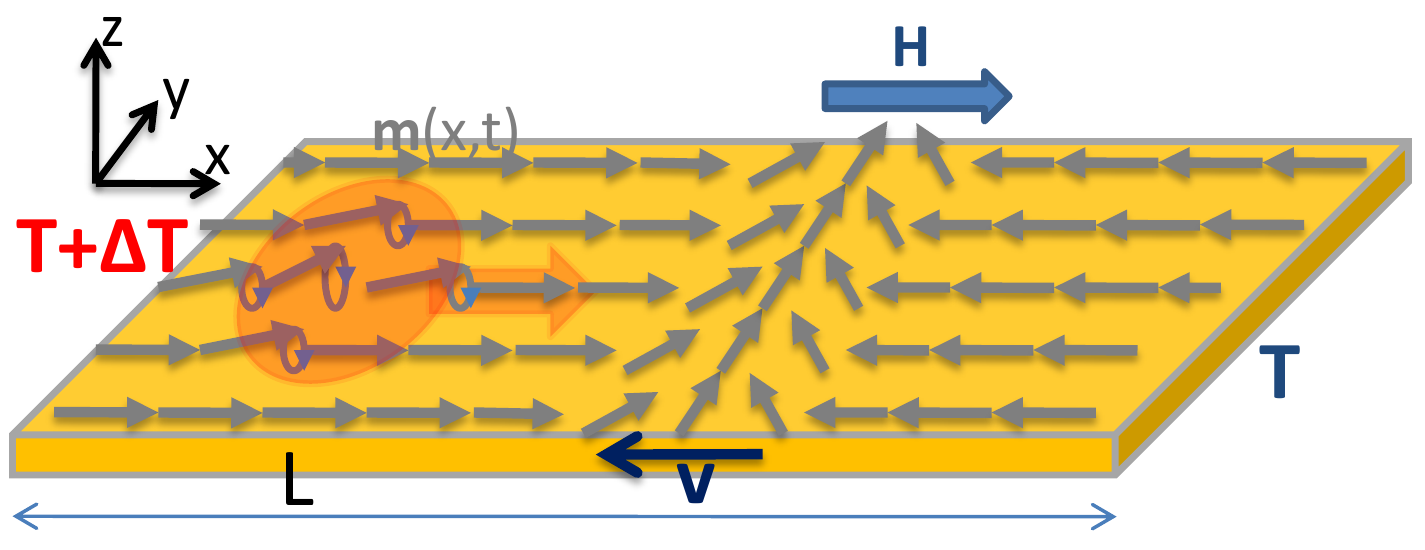}
\caption{The magnon current induced by temperature gradient
exerts spin torque on the magnetization, according to the conservation of angular momentum,
which can lead to a domain-wall motion. The inverse effect of
magnon current induced by magnetization motion is also possible,
wherein the collective magnetic texture $\mathbf{m}(x,t)$ is controlled by
effective field $\mathbf{H}$.}
\label{System} 
\end{figure}

\section{Phenomenological description of magnonic and thermal currents
in textured magnets}
We construct a general phenomenological description
of magnonic and thermal currents in textured magnets (Fig. \ref{System})
based on several thermodynamic variables such as the direction of
the slow (averaged over the magnonic excitations) spin density
$\mathbf{m}_{s}$ (the label $s$ is dropped in this section), density
of magnons $\rho$ and density of energy $\rho_{U}$. The ferromagnet
can be taken out of equilibrium by applying temperature, chemical
potential and magnetization gradients, while the equilibrium state
can be topologically nontrivial, e.g., a magnetic DW or vortex. In the absence of bias, the
ferromagnet would then evolve back towards equilibrium according to the
equations of motion. Following the standard route, we identify thermodynamic
variables $x_{k}$ and their conjugates (generalized forces) $X_{k}=\partial\mathbb{S}/\partial x_{k}$
by writing the entropy $\mathbb{S}$ production in the following form: 
\begin{equation}
\dot{\mathbb{S}}=\sum X_{k}\dot{x}_{k}\,.
\end{equation}
 The local conservation laws of energy and (approximately) magnon number provide
us with continuity relations: $\dot{\rho}=-\boldsymbol{\nabla}\mathbf{j}-(\rho-\rho_0)/\tau_{\alpha}$
and $\dot{\rho}_{U}=-\boldsymbol{\nabla}\mathbf{j}_{U}$, where we
introduced the magnon-number ($\mathbf{j}$) and energy ($\mathbf{j}_{U}$)
current densities, and $\tau_{\alpha}$ corresponds to the life time of magnons (that are in excess of a local-equilibrium value $\rho_0$).
We now write the rate of the entropy production in a standard manner
\cite{Landau:1984}: 
\begin{equation}
\dot{\mathbb{S}}=-\int d^{3}\mathbf{r}\,\dfrac{\boldsymbol{\nabla}\mathbf{j}_{U}+\mu\dot{\rho}+\boldsymbol{\mathcal{H}}\cdot\mathbf{\dot{m}}}{T}\:,\label{Entropy_Rate}
\end{equation}
 where the conjugate force corresponding to the magnetic (spin-density) direction $\mathbf{m}$
is defined as $-\delta_{\mathbf{m}}\mathbb{S}|_{\mathbf{j}_{U}(\mathbf{j})=0}=\boldsymbol{\mathcal{H}}/T$.
By straightforward manipulation, in which we introduce the modified
energy current $\mathbf{j}_{q}=\mathbf{j}_{U}-\mu\mathbf{j}$, we
arrive at the following equation for the rate of the entropy production:
\begin{equation}
\dot{\mathbb{S}}=\int d^{3}\mathbf{r}\,\left[\boldsymbol{\nabla}\left(\dfrac{1}{T}\right)\mathbf{j}_{q}-\dfrac{\boldsymbol{\nabla}\mu}{T}\mathbf{j}-\dfrac{\boldsymbol{\mathcal{H}}}{T}\cdot\mathbf{\dot{m}}\right]\,,\label{Entropy_Rate_1}
\end{equation}
 where in Eq. (\ref{Entropy_Rate}) we integrated the term involving
$\mathbf{j}_{q}$ by parts, used the continuity equations and disregarded
magnon decay $\propto(\rho-\rho_0)/\tau_{\alpha}$. The latter is justified when
either (i) the number of magnons is (to a good approximation) conserved, which means that the relevant size of the
system is smaller than the magnon decay length corresponding to $\tau_{\alpha}$, or (ii) no build-up of magnons takes place due to fast relaxation or uniform current generation (thus $\rho\approx\rho_0$).
The forces conjugate to the fluxes can be immediately
identified as $-\partial_{\mathbf{j}_{q}}\dot{\mathbb{S}}|_{\mathbf{m},\mathbf{j}=0}=-\boldsymbol{\nabla}\left(1/T\right)$
and $-\partial_{\mathbf{j}}\dot{\mathbb{S}}|_{\mathbf{m},\mathbf{j}_{q}=0}=\boldsymbol{\nabla}\mu/T$.
Formally, Eq. (\ref{Entropy_Rate_1}) is identical to the one used in Ref. \cite{Kovalev:Sep2009},
which suggests similarities between phenomenological theories for magnons
and electrons.

We now relate the currents $\mathbf{j}$ and $\mathbf{j}_{q}$ as
well as the time derivative of the collective spin-density direction,
$\mathbf{\dot{m}}$, to the thermodynamic conjugates via kinetic coefficients.
The kinetic coefficients can be further identified by noting that
the currents $\mathbf{j}$ and $\mathbf{j}_{q}$ are determined by
the chemical potential and temperature gradients as well as the magnetic
texture dynamics, which exerts fictitious Berry phase gauge fields
on the magnons. By assuming the spin-rotational symmetry of the magnetic
texture and isotropy in real space, we obtain the magnon and energy current
gradient expansion: 
\begin{align}
-\partial_{i}\mu=&\Upsilon_{ik}j_{k}+\Pi_{ik}\partial_{k}T/T-p\left(\mathbf{m}\times\partial_{i}\mathbf{m}+\beta\partial_{i}\mathbf{m}\right)\cdot\mathbf{\dot{m}}\,,\nonumber\\
(j_{q}^{h})_{i}=&\Pi_{ik}^{T}j_{k}-\kappa_{ik}\partial_{k}T-p_{1}\left(\mathbf{m}\times\partial_{i}\mathbf{m}+\beta_{1}\partial_{i}\mathbf{m}\right)\cdot\mathbf{\dot{m}}\,,\nonumber\\
&\hspace{-1cm}\mathfrak{s}(1+\alpha\mathbf{m}\times)\mathbf{\dot{m}}+\mathbf{m}\times\mathbf{H}_{\rm eff}=p\left[\partial_{i}\mathbf{m}+\beta\mathbf{m}\times\partial_{i}\mathbf{m}\right]j_{i}\nonumber\\
&+p_{1}\left[\partial_{i}\mathbf{m}+\beta_{1}\mathbf{m}\times\partial_{i}\mathbf{m}\right]\partial_{i}T/T\,,
\label{ohmD-1}
\end{align}
where the equation for the magnon current has been inverted, $\Upsilon_{ik}$,
$\Pi_{ik}$ and $\kappa_{ik}$ are the resistivity, Peltier and thermal
conductivity tensors, respectively, which are in general temperature
and texture dependent ($p_{1}\beta_{1}$ has to be treated as
one coefficient when $p_{1}=0$) and the LLG part has been completed
by employing the Onsager reciprocity principle. The heat current in
Eq. (\ref{Entropy_Rate_1}) includes the contribution from the magnetic
energy contained in the texture $\mathbf{j}_{q}^{t}$ (e.g., exchange
energy flow accompanying the DW motion). In situations when the texture
energy contribution can be separated from the ``real'' heat current,
we choose to subtract this contribution, i.e., $\mathbf{j}_{q}^{h}=\mathbf{j}_{q}-\mathbf{j}_{q}^{t}$
in Eq. (\ref{ohmD-1}). By invoking the time reversal argument, one
can show that the gradient expansion of the texture energy contribution
has the form $(j_{q}^{t})_{i}=p^{t}\beta^{t}\partial_{i}\mathbf{m}\cdot\mathbf{\dot{m}}$.
Equation (\ref{Entropy_Rate_1}) can now be reformulated using the
{}``pure'' heat current $\mathbf{j}_{q}^{h}=\mathbf{j}_{q}-\mathbf{j}_{q}^{t}$
where as it follows from Eq. (\ref{Entropy_Rate_1}) the effective
field has to be redefined as $\mathbf{H}_{\rm eff}=\boldsymbol{\mathcal{H}}-p^{t}\beta^{t}\partial_{i}\mathbf{m}\partial_{i}\left(1/T\right)$.
The new notations allow for more natural separation of the magnetic
and heat degrees of freedom, e.g., in the simplest approximation one
can assume that even in an out-of-equilibrium situation, when $\partial_{i}T\neq0$
and $\partial_{i}\mu\neq0$, $\mathbf{H}_{\rm eff}$ depends only
on the instantaneous texture $\mathbf{m}(\mathbf{r},t)$. In general,
we may expand $\mathbf{H}_{\rm eff}$ phenomenologically in terms
of small $\partial_{i}T$ and $\partial_{i}\mu$. The form of Eqs.
(\ref{ohmD-1}) is the same for both sets of variables ($\mathbf{j}_{q}^{h}\leftrightarrow\mathbf{j}_{q}$,
$\mathbf{H}_{\rm eff}\leftrightarrow\boldsymbol{\mathcal{H}}$)
but the kinetic coefficients are different. The rest of the paper
relies on equations written for ($\mathbf{j}_{q}^{h}$, $\mathbf{H}_{\rm eff}$).
Note that in Ref. \cite{Kovalev:Sep2009} it was implied that the
latter set of variables had been used.

The tensors $\Upsilon_{ik}$, $\Pi_{ik}$ and $\kappa_{ik}$ can depend
on temperature and texture, i.e., to the leading order, as 
\begin{align}
\kappa_{ik}(\Pi_{ik},\Upsilon_{ik})=&\delta_{ik}\left[\kappa(\Pi,\Upsilon)+\eta_{\kappa(\Pi,\Upsilon)}(\partial_{l}\mathbf{m})^{2}\right] \nonumber\\
&\hspace{-1cm}+\eta_{\kappa(\Pi,\Upsilon)}^{'}\partial_{i}\mathbf{m}\cdot\partial_{k}\mathbf{m}+b_{\kappa(\Pi,\Upsilon)}\mathbf{m}\cdot(\partial_{i}\mathbf{m}\times\partial_{k}\mathbf{m})\,.\nonumber
\end{align}
Equations (\ref{ohmD-1}) should also be applicable to magnetization
dynamics in the presence of charge currents (the magnon current has
to be replaced by the charge current) as discussed in Ref. \cite{Kovalev:Sep2009}.
These equations should describe such fictitious-field induced effects
on magnons (electrons) as the Hall effect (coefficient $b_{\Upsilon}$),
the Ettingshausen effect (coefficient $b_{\Pi}/\kappa$), the Nernst
effect (coefficient $b_{\Pi}/T$), and the Righi-Leduc effect (coefficient
$b_{\Pi}/\Pi$).

The gradient expansion in Eq.~\eqref{ohmD-1} assumes short magnon wavelength compared to the characteristic textures length scale. In YIG, for example, the former is $\sim1$~nm at room temperature, which should not pose a serious constraint for such adiabatic description.

\section{Bottom up construction of phenomenology for thermal magnons}
Since we concentrate on the low temperature limit (on the scale set by the Curie temperature) the LLG phenomenology remains reliable microscopically. Coarse graining thermal fluctuations thereof would however lead to modified effective quantities (such as spin density and stiffness constant). Consider a ferromagnet with space- and time-dependent spin density
$s\mathbf{m}(\mathbf{r},t)$ with magnitude $s$ saturated at a constant value
and direction described by a unit vector $\mathbf{m}(\mathbf{r},t)$. The
spin density is assumed to have two components -- fast and slow where
the slow component slowly varies in space and time with much larger
characteristic scales compared to the fast component. The effect of
topological gauge fields due to magnetic textures of the slow component
can be captured by considering the Lagrangian \cite{Guslienko:Jan2010}:
\begin{equation}
\mathcal{L}=\int d^{3}\mathbf{r}\left[\mathbf{D}(\mathbf{m})\cdot\mathbf{\dot{m}}-E(\mathbf{m},\partial_{\alpha}\mathbf{m})\right]\:,\label{Lagrangian}
\end{equation}
 where $\mathbf{D}(\mathbf{m})=s[\mathbf{n}\times\mathbf{m}]/(1+\mathbf{m}\cdot\mathbf{n})$
is the vector potential of the Wess-Zumino action with an arbitrary
$\mathbf{n}$ pointing along the Dirac string, $E(\mathbf{m},\partial_{\alpha}\mathbf{m})$
is the magnetic energy density describing the exchange energy as well
as the external and anisotropy fields. Equation (\ref{Lagrangian})
leads to the Landau-Lifshitz (LL) equation $s\mathbf{\dot{m}}-\mathbf{m}\times\delta_{\mathbf{m}}E=0$.
We assume in our description that the magnetic energy density is given
by $E/M_{s}=A(\partial_{\alpha}\mathbf{m})^{2}-\mathbf{m}\cdot\mathbf{H}_{m}/2-\mathbf{m}\cdot\mathbf{H}$
where $A$ is the exchange stiffness, $M_{s}$ is the saturation magnetization,
$\mathbf{H}_{m}$ describes magnetostatic and magnetocrystalline anisotropies
(largely ignored in our description of thermal magnons at sufficiently
high temperatures; but would otherwise suppress the fictitious forces acting on magnons \cite{Dugaev:jul2005}), and $\mathbf{H}$ is the external magnetic field.
For the purpose of deriving the equations of motion, we use a coordinate transformation after which the $z-$axis points
along the spin density of the slow dynamics. In the new coordinate
system, small excitations will only have $m_{x}$ and $m_{y}$ components.

Equations for the slow dynamics in the original (lab) frame can be obtained from the LL equation with a dissipative term 
by coarse-graining over fast variables, i.e., $s\left\langle \mathbf{m}\right\rangle_{\rm fast}=\mathfrak{s}(\mathbf{r},t)\mathbf{m}_{s}$
with the final result: 
\begin{align}
\mathfrak{s}\mathbf{\dot{m}}_{s}+\mathbf{m}_{s}\times\mathbf{H}_{\rm eff}^{s}=&\hbar\partial_{\alpha}\mathbf{j}_{\alpha}+\alpha\mathbf{m}_{s}\times(\mathbf{m}_{s}\times\mathbf{H}_{\rm eff}^{s})\nonumber\\
 & -\hbar(\alpha-\beta)\mathbf{m}_{s}\times(\partial_{\alpha}\mathbf{j}_{\alpha})\:,
\label{LL}
\end{align}
 where the l.h.s. is written for the magnon-averaged spin density $\mathfrak{s}$,
$\mathbf{m}$ and $\mathbf{m}_{s}$ are unit vectors, i.e., $\mathbf{m}_{f}=\mathbf{m}-(\mathfrak{s}/s)\mathbf{m}_{s}$
corresponds to the fast dynamics, the spin current density $\hbar\mathbf{j}_{\alpha}=M_{s} A\left[\mathbf{m}_{f}\times\partial_{\alpha}\mathbf{m}_{f}\right]$, $\mathbf{H}_{\rm eff}^{s}=-\delta_{\mathbf{m}_{s}}E\{(\mathfrak{s}/s)\mathbf{m}_{s},\partial_{\alpha}[(\mathfrak{s}/s)\mathbf{m}_{s}]\}$ is the effective field of the slow dynamics
and $\beta$ accounts for the fact that magnons misalign with the
direction of the slow dynamics \cite{Zhang:sep2004}. Note that Eq. (\ref{LL}) assumes short magnon wavelengths compared to the slow texture length scale. By multiplying
Eq. (\ref{LL}) by $1+\alpha\mathbf{m}_{s}\times$ from the left and
using an approximation $\partial_{\alpha}\mathbf{j}_{\alpha}\approx(j_{i}\partial_{i})\mathbf{m}_{s}$
($j_{i}$ corresponds to the magnon number flux) which is true for
slowly varying textures, we recover the Landau-Lifshitz-Gilbert (LLG)
equation: 
\begin{equation}
\mathfrak{s}(1+\alpha\mathbf{m}_{s}\times)\mathbf{\dot{m}}_{s}+\mathbf{m}_{s}\times\mathbf{H}_{\rm eff}^{s}=\hbar\left[1+\beta\mathbf{m}_{s}\times\right](j_{i}\partial_{i})\mathbf{m}_{s}\:,\label{LLG}
\end{equation}

In order to describe small excitations (spin waves) in the coordinates with the $z-$axis pointing along the slow dynamics, we introduce
$3\times3$ rotation matrix $\hat{R}=\exp(\psi\hat{J}_{z})\exp(\theta\hat{J}_{y})\exp(\phi\hat{J}_{z})$
with $\hat{J}_{\alpha}$ being the $3\times3$ matrix describing infinitesimal
rotation along the axis with index $\alpha$. In the new coordinates,
we have $\mathbf{m}\rightarrow\mathbf{m}^{'}=\hat{R}\mathbf{m}$ and
$\partial_{\mu}\rightarrow(\partial_{\mu}-\hat{\mathcal{A}}_{\mu})$
with $\hat{\mathcal{A}}_{\mu}=(\partial_{\mu}\hat{R})\hat{R}^{-1}$
(the index $\mu=0,..,3$ denotes the time and space coordinates).
Since the matrix $\hat{\mathcal{A}}_{\mu}$ is skew-symmetric, we
can introduce a vector $\mathbf{\boldsymbol{\mathcal{A}}}_{\mu}$
so that $\hat{\mathcal{A}}_{\mu}\mathbf{m}=\mathbf{\boldsymbol{\mathcal{A}}}_{\mu}\times\mathbf{m}$.
In a specific gauge with the Euler angle $\psi=0$, the elements of $\mathbf{\boldsymbol{\mathcal{A}}}_{\mu}$
become $\mathbf{\boldsymbol{\mathcal{A}}}_{\mu}=(-\sin\theta\partial_{\mu}\phi,\partial_{\mu}\theta,\cos\theta\partial_{\mu}\phi)$.
The equation describing spin waves follows from the LL equation subject
to the coordinate transformation: 
\begin{equation}
i(\partial_{t}-i\mathcal{A}_{0}^{z})m_{+}=A\left(\partial_{\alpha}/i-\mathcal{A}_{\alpha}^{z}\right)^{2}m_{+}+V(\mathbf{r})m_{+}\:.\label{Magnons}
\end{equation}
 Here $V(\mathbf{r})=\mathbf{m}_{s}\cdot\mathbf{H}/s-A(\mathbf{\boldsymbol{\mathcal{A}}}_{x}^{2}+\mathbf{\boldsymbol{\mathcal{A}}}_{y}^{2})/2$
is the effective potential and $\mathbf{m}_{s}$
is the unit vector along the spin density of the slow dynamics. Second order terms in the effective potential $\sim \mathbf{\boldsymbol{\mathcal{A}}}_{\mu}^2$ are not treated systematically as there are similar corrections that lead to the coupling between
the circular components of spin wave $m_{\pm}=m_{x}^{'}(\mathbf{r},t)\pm im_{y}^{'}(\mathbf{r},t)$ where $m_{x(y)}^{'}(\mathbf{r},t)$ describes the transverse
excitations in the transformed coordinates with the $z$ axis pointing
along the direction of $\mathbf{m}_{s}$ (in the absence of texture
$m_{+}\sim\exp(i\mathbf{q}\mathbf{r}+i\omega_{q}t)$). We
omitted damping terms for the case when the region of interest is smaller than the
length corresponding to the lifetime of spin waves. The coupling between
the circular components of spin wave due to anisotropies 
is disregarded since we assume that the exchange effects dominate. By quantizing
spin waves, we introduce the field operator $\psi=\sqrt{s/2\hbar^{2}}m_{+}=\sqrt{1/\hbar}{\textstyle \sum_{\mathbf{q}}}b_{\mathbf{q}}e^{i\mathbf{q}\mathbf{r}}$
corresponding to $m_{+}$ and describing magnons with spectrum $\omega_{q}=V(\mathbf{r})+Aq^{2}$
where $b_{\mathbf{q}}^{\dagger}$ and $b_{\mathbf{q}}$ are creation
and annihilation operators. In such notations, the current is written
as $j_{\alpha}=-i\hbar^{2}A(\psi^{\dagger}\partial_{\alpha}\psi-\psi\partial_{\alpha}\psi^{\dagger})$.
Note that Eq. (\ref{Magnons}) describes charged particles moving
in the fictitious electric $\mathcal{E}_{\alpha}=-\partial_{t}\mathcal{A}_{\alpha}^{z}-\partial_{\alpha}(\mathcal{A}_{0}^{z}+V(\mathbf{r}))=\hbar\mathbf{\widetilde{m}}_{s}\cdot(\partial_{t}\mathbf{\widetilde{m}}_{s}\times\partial_{\alpha}\mathbf{\widetilde{m}}_{s})-\hbar\partial_{\alpha}V(\mathbf{r})$
and magnetic $\mathcal{B}_{i}=(\hbar/2)\epsilon^{ijk}\mathbf{\widetilde{m}}_{s}\cdot(\partial_{k}\mathbf{\widetilde{m}}_{s}\times\partial_{j}\mathbf{\widetilde{m}}_{s})$
fields produced by the magnetic texture. In the diffusive regime,
the transport of magnons due to such fields can be found by solving
the Boltzmann equation within the relaxation-time approximation.

The texture independent transport coefficients can be expressed through
the following integrals: 
\begin{equation}
\mathcal{J}_{n}^{\alpha\beta}=\dfrac{1}{(2\pi)^{3}\hbar}\int d\varepsilon\tau(\varepsilon)(\varepsilon-\mu)^{n}\left(-\dfrac{\partial f_{0}}{\partial\varepsilon}\right)\int dS_{\varepsilon}\dfrac{\upsilon_{\alpha}\upsilon_{\beta}}{|\boldsymbol{\upsilon}|}\:,\label{Boltzmann}
\end{equation}
 where $\mu$ is the chemical potential of magnons, $\tau(\varepsilon)$
is the relaxation time, $\varepsilon(\mathbf{q})=\hbar\omega_{q}$,
$\upsilon_{\alpha}=\partial\omega_{q}/\partial q_{\alpha}$, $dS_{\varepsilon}$
is the area $d^{2}q$ corresponding to the constant energy $\varepsilon(\mathbf{q})=\varepsilon$
and $f_{0}=\left\{ \exp\left[(\varepsilon-\mu)/k_{B}T\right]-1\right\} ^{-1}$
is the Bose-Einstein equilibrium distribution. Thus the resistivity
tensor is $\rho=\mathcal{J}_{0}^{-1}$, the analogue of the Peltier
coefficient for magnons is $\Pi=-\mathcal{J}_{1}\mathcal{J}_{0}^{-1}$,
and the thermal conductivity is $\kappa=(\mathcal{J}_{2}-\mathcal{J}_{1}\mathcal{J}_{0}^{-1}\mathcal{J}_{1})/T$.
The relaxation time $\tau$ contains the conserving number of magnons
part $\tau_{c}$, e.g., corresponding to magnon-magnon interactions and disorder scattering,
and the nonconserving part, e.g., corresponding to the Gilbert damping
$\tau_{\alpha}$. We can sum up different contributions with
different dependence on temperature according to $1/\tau=1/\tau_{c}+1/\tau_{\alpha}$.
In this work, we assume that the scattering is dominated by the Gilbert
damping contribution $\tau_{\alpha}\sim(2\alpha\omega)^{-1}$, where
$\alpha$ is the Gilbert damping, which is consistent with our bottom
up approach (in some cases, however, the contribution corresponding
to the temperature independent scattering length gives better agreement
to the experiment \cite{Douglass:1963}). By taking the quadratic
spectrum for YIG at room temperature and $\alpha\sim10^{-3}$ \cite{Heinrich:aug2011},
we obtain the mean free path $\sim100\:\mbox{nm}$ from the $\tau_{\alpha}$
relaxation time for magnons. 

The above description of magnons contains dissipative
$\beta$ corrections \cite{Zhang:sep2004} related to a small magnon
mistracking of the slow magnetic texture. Such corrections have been extensively
studied for charge flows in transition metals and one can use the
analogy between the electronic and magnonic systems in order to introduce
such corrections into the equations for currents. By following this
prescription, one can arrive at Eqs. (\ref{ohmD-1}) and (\ref{LLG}) with the transport
coefficients given by the Boltzmann equation. The coefficient $\beta$
can then be estimated as $\beta\sim\alpha$ based on the absence of
any additional energy scales, apart from the thermal energy, in the low-temperature (compared to the Curie temperature) limit captured by our
bottom up approach. More careful estimates of $\beta$ can be obtained
microscopically by solving equation for nonequilibrium spin polarization
in the presence of the transverse relaxation \cite{Tserkovnyak:Oct2006}
or by using the scattering matrix approach \cite{Bauer:jan2010}. 

By relating the microscopic description based on Eq. (\ref{LLG})
with Eqs. (\ref{ohmD-1}), we can immediately identify some of the
kinetic coefficients in Eq. (\ref{ohmD-1}), i.e., $p_{1}=0$, $p_{1}\beta_{1}=0$,
$p=-\hbar$ and $\alpha/\beta\sim1$. Note that $p_{1}$ and $p_{1}\beta_{1}$
can be nonzero in systems with more than one magnon band and/or added anisotropies. In order
to find the remaining coefficients, one could use Boltzmann equation
and the relaxation time approximation applied to three dimensional
magnons with Bose distribution [e.g., Eq. (\ref{Boltzmann})]. Such
description can be further improved by considering the dissipative
corrections described by $\eta$ terms.

\begin{figure}
\onefigure[width=0.9\linewidth]{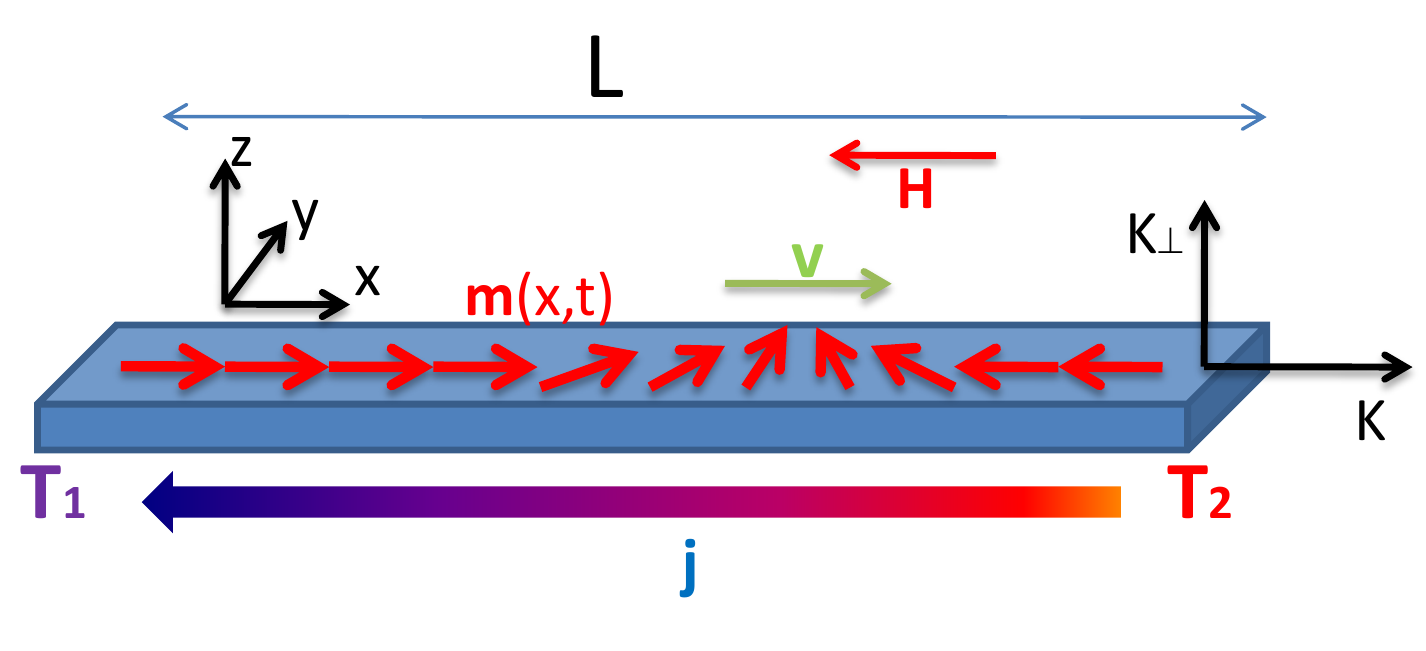}
\caption{(Color online) A domain-wall moves towards the hotter end of the wire
due to the flow of magnons $j$. Here, we consider transverse head-to-head
N{\'e}el domain wall parallel to the $y$ axis in the easy $xy$ plane.
The constants $K$ and $K_{\perp}$ describe the easy axis and easy
plane anisotropy.}
\label{DomainWall} 
\end{figure}

\section{Magnonic domain wall motion by temperature gradient} 
In this section we show how Eqs. (\ref{ohmD-1}) can be used to describe
the domain wall motion by employing the Walker ansatz. To this end,
we will describe the domain wall in Fig. \ref{DomainWall} by the
Walker ansatz valid for weak field and current/temperature biases
\cite{Tserkovnyak:apr2008a,Kovalev:nov2010}:
\begin{equation}
\varphi(\mathbf{r},t)\equiv\Phi(t),\quad\ln\tan\dfrac{\theta(\mathbf{r},t)}{2}\equiv\dfrac{x-X(t)}{W(t)}\:,\label{Walker}
\end{equation}
where the position-dependent spherical angles $\varphi$ and $\theta$
parametrize the magnetic configuration as $\mathbf{m}(x)=m(x)(\cos\theta,\sin\theta\cos\varphi,\sin\theta\sin\varphi)$,
$X(t)$ parametrizes the net displacement of the wall along the $x$
axis, and we assume that the driving forces ($H$, $j$ and $j_{q}$)
are not too strong so that the wall preserves its shape and only its
width $W(t)$ and out-of-plane tilt angle $\Phi(t)$ undergo small
changes. By substituting the ansatz (\ref{Walker}) in Eq. (\ref{LLG})
with the effective field given by $\mathbf{H}_{\rm eff}=\gamma \mathfrak{s}\left[(H+(\mathfrak{s}/s)Km_{x})\mathbf{x}-(\mathfrak{s}/s)K_{\perp}m_{z}\mathbf{z}+(\mathfrak{s}/s)A\nabla^2\mathbf{m}\right]$,
we obtain: 
\begin{align}
\dot{\Phi}+\dfrac{\alpha\dot{X}}{W}&=\gamma H-\dfrac{p\beta j}{\mathfrak{s}W}\:,\nonumber\\
\dfrac{\dot{X}}{W}-\alpha\dot{\Phi}&=\dfrac{\gamma (\mathfrak{s}/s)K_{\perp}\sin2\Phi}{2}-\dfrac{pj}{\mathfrak{s}W}\:,\nonumber\\
W&=\sqrt{\dfrac{A}{K+K_{\perp}\sin^{2}\Phi}}\:,
\label{Walker1}
\end{align}
where $\gamma$ is the gyromagnetic ratio, $A$ is the stiffness constant, and $K$ and $K_{\perp}$ describe
the easy axis and easy plane anisotropies, respectively. Note that the factor $\mathfrak{s}/s$ before the stiffness constant appears due to coarse-graining over fast variables in Eq. (\ref{LL}). This dependence of the effective stiffness constant on the average spin density can lead to additional contribution to the domain wall velocity \cite{Hinzke:Jul2011}. In this work, we concentrate on the low temperature limit (on the scale set by the Curie temperature) in which case $\mathfrak{s}\sim s$ and the domain wall motion is dominated by the spin-transfer torque contribution. The steady
state solution of Eq. (\ref{Walker1}) below the Walker breakdown
with $\Phi(t)=\mbox{const}$ and $X=\upsilon t$ leads to the result
$\upsilon=(\gamma HW+\hbar\beta j/s)/\alpha$ or equivalently 
\begin{equation}
\upsilon=\dfrac{\gamma W}{\alpha}H+\dfrac{F_{0}\beta}{6\pi^{2}\lambda s\alpha}\partial_{x}T\:,\label{vel1}
\end{equation}
where $\lambda$ is the thermal magnon wavelength and we introduce
a numerical dimensionless factor $F_{0}(x)=\int d\epsilon\epsilon^{3/2}e^{\epsilon+x}/(e^{\epsilon+x}-1)^{2}\sim1$ evaluated at the magnon gap, $x=\hbar\omega_0/k_BT$, which corresponds to dimesionless part of the integral $\mathcal{J}_1$ in Eq. (\ref{Boltzmann}).
From Eq. (\ref{vel1}), we estimate the domain wall velocity in YIG
at room temperature to be $1\:\mbox{cm}/\mbox{sec}$ for the temperature
gradient $1\:\mbox{K}/\mu\mbox{m}$.

\section{Peltier effect in quasi-one-dimensional wire}  
Peltier effect describes the heat flow accompanying the flow of carriers.
Since the moving magnetic texture will induce the flow of magnons,
one can discuss the heat currents resulting from such a process. In
this section, we concentrate on quasi-one-dimensional systems with
a DW propagating in a one dimensional wire connected to two reservoirs
in quasi-equilibrium state (Fig. \ref{DomainWall}). Contrary to the
previous sections, the wire can be in the ballistic as well as in
the diffusive regime. By writing the equation for the entropy production
in the form analogous to the microscopic form in Eq. (\ref{Entropy_Rate_1}):
\begin{equation}
\dot{\mathbb{S}}=\dfrac{L}{T}\left[-j_{q}\dfrac{T_{2}-T_{1}}{TL}-j\dfrac{\mu_{2}-\mu_{1}}{L}-\dot{X}\dfrac{2MH}{L}\right]\,,\label{Entropy_Rate2}
\end{equation}
and identifying the thermodynamics variables $j_{q}$, $j$ and $\dot{X}$,
we can phenomenologically generalize Eq. (\ref{ohmD-1}) to systems
possessing domain wall (solitonic) solutions describable by a single
generalized coordinate $X$. The general phenomenological equations
for the domain wall dynamics become:
\begin{align}
\dot{X}&=-\mathcal{O}_{X}(2M_{s}H/L)-\mathcal{O}_{Xj}\mathcal{E}-\mathcal{O}_{XT}j_{q}\:,\nonumber\\
\mathcal{E}_{T}/T&=\mathcal{O}_{T}j_{q}-\mathcal{O}_{Tj}\mathcal{E}-\mathcal{O}_{XT}(2M_{s}H/L)\:,\nonumber\\
j&=\mathcal{O}_{j}\mathcal{E}+\mathcal{O}_{Tj}j_{q}+\mathcal{O}_{Xj}(2M_{s}H/L)\:,
\label{Onsager}
\end{align}
 where $L$ is the length of the wire, $\mathcal{E}=-(\mu_{2}-\mu_{1})/L$,
$\mathcal{E}_{T}=-(T_{2}-T_{1})/L$ and the kinetic coefficients now
correspond to the whole wire, i.e., $\mathcal{O}_{T}=1/(\kappa+\kappa ZT)$,
$\mathcal{O}_{j}=1/(\Upsilon+\Upsilon ZT)$ and $\mathcal{O}_{Tj}=1/(\Pi+\Pi/ZT)$
with the conventional figure of merit relation for magnons $ZT=\Pi^{2}/\Upsilon T\kappa$.
Using Eq. (\ref{Onsager}) one can fully describe the interplay between
the domain wall motion and magnon/heat currents in the quasi-one-dimensional
systems. The kinetic coefficients can be extracted from the bulk values
in the diffusive regime (see below) or can be calculated by scattering
theory methods in the ballistic regime \cite{Bauer:jan2010}. 

In the following, we analyze the efficiency of heat pumping by a moving
domain wall in a system depicted in Fig. \ref{DomainWall}. Given
that the rate of dissipation due to the DW motion in Fig. \ref{DomainWall}
is $2M_{s}H\dot{X}$ and it can be assumed to be divided equally between
reservoirs, we can calculate the ratio between the useful heat taken
from the cooled reservoir and the dissipated heat as the domain wall
moves from the left end of the wire to the right end. Such ratios
are often calculated in order to characterize thermoelectric circuits
\cite{Mahan:1997}. By maximizing the rate of cooling as a function
of the domain wall velocity we obtain: 
\begin{equation}
COP_{\rm cool}=\dfrac{j_{q}^{\rm cold}}{2HM\dot{X}}=\dfrac{T_{c}}{T_{h}-T_{c}}\dfrac{\sqrt{1+TZ_{\rm mc}}-T_{h}/T_{c}}{\sqrt{1+TZ_{\rm mc}}+1}\:.\label{MaxCool}
\end{equation}
 Here $j_{q}^{\rm cold}$ is the heat current leaving the cooled
reservoir and we define the magnetocaloritronic figure of merit \cite{Kovalev:Sep2009}
by analogy to the thermoelectric figure of merit $ZT$: 
\begin{equation}
TZ_{\rm mc}=\dfrac{\mathcal{O}_{XT}^{2}}{\mathcal{O}_{T}\mathcal{O}_{X}}\:,\label{ZT}
\end{equation}
where we assume that $\mathcal{E}=0$ and such definition of $TZ_{\rm mc}$
ensures that Eq. (\ref{MaxCool}) is identical to the expression for
the thermolelectric $COP_{\rm cool}$ after $TZ_{\rm mc}$ is
replaced by $ZT$. The figure of merit in Eq. (\ref{ZT}) is also
related to the maximum efficiency of the magnetocaloritronic power
generator in a device driven by ac magnetic field \cite{Kovalev:Sep2009}
contrary to the geometry optimized magnetothermopower in applied dc
magnetic field \cite{Heremans:prl2001}. By taking some particular
DW solution we can relate Eqs. (\ref{ohmD-1}) with Eqs. (\ref{Onsager}),
e.g., for a transverse head-to-head N{\'e}el DW solution in Eq. (\ref{Walker}),
$\mathcal{O}_{X}=LW/2\alpha s+\Upsilon(1+ZT)\mathcal{O}_{Xj}^{2}$,
$\mathcal{O}_{Xj}(1+ZT)=(p\beta/\Upsilon-p_{1}\beta_{1}ZT/\Pi)/\alpha s$
and $\mathcal{O}_{XT}(1+ZT)=(p\beta ZT/\Pi+p_{1}\beta_{1}/T\kappa)/\alpha s$. 

We now estimate $TZ_{\rm mc}$ with $p_{1}\beta_{1}=0$ according
to Eq. (\ref{LLG}) and assuming that the scattering is dominated
by nonconserving mechanisms described by $\tau_{\alpha}\sim(2\alpha\omega)^{-1}$.
The corresponding mean-free path is assumed to be smaller than the
system size in order to ensure the diffusive limit [in principle the
ballistic regime can also be treated by Eq. (\ref{Onsager}) in which
the kinetic coefficients should be found using an appropriate microscopic approach].
By invoking Eq. (\ref{Boltzmann}) we express the figure of merit
as follows: 
\begin{equation}
TZ_{\rm mc}\approx\dfrac{2F_1\sqrt{k_{B}T}(\beta/\alpha)^{2}}{3\pi^{2}\sqrt{A/\hbar}WLs}\sim\dfrac{(\beta/\alpha)^{2}}{W^{2}\lambda s/\hbar}\,,\label{Estimate}
\end{equation}
 where $\lambda$ is the thermal magnon
wavelength, the wire length is taken to be $L\sim W$, and we introduce a numerical dimensionless factor $F_{1}(x)=[\mathcal{J}_1^2/(\mathcal{J}_0 \mathcal{J}_2)]\int d\epsilon\epsilon^{3/2}e^{\epsilon+x}/[(e^{\epsilon+x}-1)^{2}(\epsilon+x)]$  evaluated at the magnon gap, $x=\hbar\omega_0/k_BT$, which corresponds to dimesionless part of the ratio $\mathcal{J}_1^2/\mathcal{J}_2$ of integrals in Eq. (\ref{Boltzmann}). We also assume that the domain
wall size can be estimated as $W\sim\sqrt{A/\omega_{0}}$ where $\omega_{0}$
can be, e.g., the demagnetizing energy. By taking material parameters
for YIG at room temperature \cite{Novoselov:2005} we arrive at $W\lesssim100\:\mbox{nm}$
and $TZ_{\rm mc}\sim10^{-4}(\beta/\alpha)^{2}$, which is quite low. However,
larger ratios $\beta/\alpha$ could be expected as one approaches
the Curie temperature. Materials with smaller DW size should be more
efficient in heat pumping according to Eq. (\ref{Estimate}), e.g.,
we estimate that in EuO $W\sim1\,\mbox{nm}$ and $TZ_{\rm mc}\sim10^{-3}(\beta/\alpha)^{2}$
at $\sim10\:\mbox{K}$ using the following parameters: the localized spin
$S=7/2$, the lattice spacing $a_{0}=5.1~\textrm{\AA}$ and the exchange
integral $J_{0}/k_{B}\approx1\:\mbox{K}$ \cite{Sollinger:Apr2010}
($J_{0}=a_{0}sA/4S^{2}$ for a face-centered lattice). In traditional
thermoelectrics $ZT$ plummets to zero much faster than $TZ_{\rm mc}\propto\sqrt{T}$
in Eq. (\ref{Estimate}), making magnonic heat pumps promising for
cryogenic applications. Furthermore, $TZ_{\rm mc}$ scales as $s^{-1}$
with the spin density thus the dilute magnetic systems (with sufficiently
narrow $W$) should also be suitable for such applications.

To conclude, we developed a phenomenological theory describing magnon
and heat currents and the magnetization texture dynamics. Under some
simple model assumptions, we are able to extract information about
all the phenomenological parameters from the Gilbert damping, the
exchange integral, the localized spin and the lattice spacing. The
$\beta$ viscous coupling also appears in our description and
is related to the magnon dephasing time. Our estimates show that the
viscous coupling effects between magnetization dynamics and magnon
flows can be strong in materials with low spin densities and narrow
domain walls, which can allow the magnonic manipulation of magnetization dynamics
and heat pumping. This opens new prospects for thermomagnonic devices,
e.g., thermomagnonic heat pumps and generators, that at low temperatures
could effectively compete with traditional thermoelectrics. When resubmitting our manuscript, we became aware of two recent works  \cite{Hinzke:Jul2011, Yan:Oct2011} that 
numerically study the magnonic spin-transfer torque and domain wall motion and arrive at results that are consistent with our studies.

\acknowledgments
We thank Joseph Heremans and Gerrit E. W. Bauer for useful discussions. This work was supported in part by the Alfred P. Sloan Foundation, DARPA, and NSF under Grant No. DMR-0840965.


\begin{thebibliography}{10}
\expandafter\ifx\csname url\endcsname\relax\def\url#1{\texttt{#1}}\fi

\bibitem{Volovik:mar1987}
\Name{{Volovik} G.~E.} \REVIEW{J. Phys. C}{20}{1987}{L83}.

\bibitem{Tatara:nov2008}
\Name{Tatara G., Kohno H. \and Shibata J.} \REVIEW{Phys. Rep.}{468}{2008}{213}.

\bibitem{Dery:may2007}
\Name{Dery H., Dalal P., Cywinski L. \and Sham L.~J.} \REVIEW{Nature}{447}{2007}{573};
\Name{{Dery} H., {Wu} H., {Ciftcioglu} B., {Huang} M., {Song} Y., {Kawakami}
  R., {Shi} J., {Krivorotov} I., {Zutic} I. \and {Sham} L.~J.} arXiv:1101.1497.

\bibitem{Zhang:sep2004}
\Name{{Zhang} S. \and {Li} Z.} \REVIEW{Phys. Rev. Lett.}{93}{2004}{127204}.

\bibitem{Thiaville:feb2005}
\Name{Thiaville A., Nakatani Y., Miltat J. \and Suzuki Y.} \REVIEW{Europhys. Lett.}{69}{2005}{990}.

\bibitem{Tserkovnyak:Oct2006}
\Name{Tserkovnyak Y., Skadsem H.~J., Brataas A. \and Bauer G. E.~W.}
  \REVIEW{Phys. Rev. B}{74}{2006}{144405}.
  
\bibitem{Kohno:nov2006}
\Name{Kohno H., Tatara G. \and Shibata J.} \REVIEW{J. Phys. Soc. Jpn.}{75}{2006}{113706}.

\bibitem{Duine:jun2007}
\Name{Duine R.~A., Nunez A.~S., Sinova J. \and MacDonald A.~H.} \REVIEW{Phys. Rev. B}{75}{2007}{214420}.

\bibitem{Kovalev:Sep2009}
\Name{Kovalev A.~A. \and Tserkovnyak Y.} \REVIEW{Phys. Rev. B}{80}{2009}{100408}.

\bibitem{Kovalev:nov2010}
\Name{Kovalev A.~A. \and Tserkovnyak Y.} \REVIEW{Solid State Commun.}{150}{2010}{500}.

\bibitem{Hals:2010}
\Name{Hals K.~M., Brataas A. \and Bauer G.~E.} \REVIEW{Solid State Commun.}{150}{2010}{461}.

\bibitem{Slonczewski:1996}
\Name{Slonczewski J.~C.} \REVIEW{J. Magn. Magn. Mater.}{159}{1996}{L1};
\Name{Berger L.} \REVIEW{Phys. Rev. B}{54}{1996}{9353}.

\bibitem{Heinrich:May2003}
\Name{Heinrich B., Tserkovnyak Y., Woltersdorf G., Brataas A., Urban R. \and
  Bauer G. E.~W.} \REVIEW{Phys. Rev. Lett.}{90}{2003}{187601}.

\bibitem{Kajiwara:mar2010}
\Name{Kajiwara Y., Harii K., Takahashi S., Ohe J., Uchida K., Mizuguchi M.,
  Umezawa H., Kawai H., Ando K., Takanashi K., Maekawa S. \and Saitoh E.}
  \REVIEW{Nature}{464}{2010}{262}.

\bibitem{Hatami:aug2007}
\Name{{Hatami} M., {Bauer} G.~E.~W., {Zhang} Q. \and {Kelly} P.~J.}
  \REVIEW{Phys. Rev. Lett.}{99}{2007}{066603};
  \Name{Yu H., Granville S., Yu D.~P. \and Ansermet J.-P.} \REVIEW{Phys. Rev. Lett.}{104}{2010}{146601}.

\bibitem{Slonczewski:Aug2010}
\Name{Slonczewski J.~C.} \REVIEW{Phys. Rev. B}{82}{2010}{054403};
\Name{{Jia} X., {Liu} K., {Xia} K. \and {Bauer} G.~E.~W.}
 arXiv:1103.3764.

\bibitem{Uchida:nov2010}
\Name{Uchida K., Xiao J., Adachi H., Ohe J., Takahashi S., Ieda J., Ota T.,
  Kajiwara Y., Umezawa H., Kawai H., Bauer G. E.~W., Maekawa S. \and Saitoh E.}
  \REVIEW{Nat Mater}{9}{2010}{894}.

\bibitem{Onose:2010}
\Name{Onose Y., Ideue T., Katsura H., Shiomi Y., Nagaosa N. \and Tokura Y.}
  \REVIEW{Science}{329}{2010}{297}.

\bibitem{Landau:1984}
\Name{Landau L. \and Lifshitz E.} \Book{Electrodynamics of Continuous Media}
  2nd Edition Vol.~8 (Pergamon, Oxford) 1984.

\bibitem{Guslienko:Jan2010}
\Name{Guslienko K.~Y., Aranda G.~R. \and Gonzalez J.~M.} \REVIEW{Phys. Rev. B}{81}{2010}{014414}.

\bibitem{Dugaev:jul2005}
\Name{Dugaev V.~K., Bruno P., Canals B. \and Lacroix C.} \REVIEW{Phys. Rev. B}{72}{2005}{024456}.

\bibitem{Douglass:1963}
\Name{Douglass R.~L.} \REVIEW{Phys. Rev.}{129}{1963}{1132}.

\bibitem{Heinrich:aug2011}
\Name{Heinrich B., Burrowes C., Montoya E., Kardasz B., Girt E., Song Y.-Y.,
  Sun Y. \and Wu M.} \REVIEW{Phys. Rev. Lett.}{107}{2011}{066604}.

\bibitem{Bauer:jan2010}
\Name{Bauer G. E.~W., Bretzel S., Brataas A. \and Tserkovnyak Y.} \REVIEW{Phys. Rev. B}{81}{2010}{024427}.

\bibitem{Tserkovnyak:apr2008a}
\Name{Tserkovnyak Y., Brataas A. \and Bauer G.~E.} \REVIEW{J. Magn. Magn. Mater.}{320}{2008}{1282}.

\bibitem{Mahan:1997}
\Name{Mahan G.} \REVIEW{Solid State Phys.}{51}{1997}{81}.

\bibitem{Heremans:prl2001}
\Name{Heremans J.~P., Thrush C.~M. \and Morelli D.~T.} \REVIEW{Phys. Rev. Lett.}{86}{2001}{2098}.

\bibitem{Novoselov:2005}
\Name{Novoselov K., Dubonos S., Morozov S., Hill E., Grigorieva I. \and Geim A.} \REVIEW{J. Low Temp. Phys.}{139}{2005}{65}.

\bibitem{Sollinger:Apr2010}
\Name{S\"ollinger W., Heiss W., Lechner R.~T., Rumpf K., Granitzer P., Krenn H.
  \and Springholz G.} \REVIEW{Phys. Rev. B}{81}{2010}{155213}.

\bibitem{Hinzke:Jul2011}
\Name{Hinzke, D. and Nowak, U.} \REVIEW{Phys. Rev. Lett.}{107}{2011}{027205}.

\bibitem{Yan:Oct2011}
\Name{Yan, P. and Wang, X. S. and Wang, X. R.} \REVIEW{Phys. Rev. Lett.}{107}{2011}{177207}.

\end{thebibliography}
\end{document}